\documentclass{PoS}

\usepackage[utf8]{inputenc}
\usepackage{lineno}

\renewcommand{\deg}{^\circ}
\newcommand{\todo}[1]{}
\renewcommand{\todo}[1]{{\color{red} TODO: {#1}}}

\newcommand{\tohide}[1]{}

\title{MAGIC observations of Dragonfly Nebula at TeV Energies using the Very Large Zenith Angle Technique}

\ShortTitle{MAGIC Observations of Dragonfly Nebula using the VLZA Technique at TeV Energies}

\author{\speaker{Darko Zarić}${}^,{}^{a}$, Razmik Mirzoyan${}^b$, Ievgen Vovk${}^b$, Petar Temnikov${}^c$, Michele Peresano${}^d$, Nikola Godinović${}^a$, Juliane van Scherpenberg${}^b$, J\"urgen Besenrieder${}^b$ for the MAGIC Collaboration\footnote{\texttt{https://magic.mpp.mpg.de/}. For collaboration list see PoS(ICRC2019)1177}\\
	\llap{$^a$} Faculty of Electrical Engineering, Mechanical Engineering and Naval Architecture\\ University of Split, Croatia\\
	\llap{$^b$} Max-Planck Institute for Physics\\ Munich, Germany\\
	\llap{$^c$} Institute for Nuclear Research and Nuclear Energy\\ Sofia, Bulgaria\\
	\llap{$^d$} CEA-Saclay-/ Irfu-DAp / UMR AIM\\
		Orme des merisiers, Bât, 709-91191 Gif-sur-Yvette Cedex, France\\
        E-mail: \email{dzaric@fesb.hr} }

\abstract{One of the brightest regions of diffuse gamma-ray emission in the northern sky is the Cygnus star-forming region, where one can assume the most energetic processes are taking place. The Dragonfly Nebula (MGRO J2019+37) is one of the brightest sources in the Cygnus region. First discovered by MILAGRO, it was later resolved into two sources by VERITAS: the faint point-like source VER J2016+371 and the bright extended source VER J2019+368. The latter accounts for the bulk of the MGRO J2019+37 emission, with the spectrum among the hardest in the TeV range.
We report the results of a dedicated MAGIC observational campaign of VER J2019+368. The data obtained with the Very Large Zenith Angle observational technique provides an effective collection area of about one square kilometer. We used $\sim$45 hours of data collected under Very Large Zenith Angles for exploring the flux of the source at TeV energies.}

\FullConference{36th International Cosmic Ray Conference -ICRC2019-\\
		July 24th - August 1st, 2019\\
		Madison, WI, U.S.A.}

\begin{document}

\section{Introduction}

The Cygnus region is a near (1.4 kpc), active star forming region and it is one of the brightest regions in diffuse gamma rays in the northern-hemisphere \cite{hunter_egret_1997}.
The Milagro Collaboration discovered a very extended source MGRO J2019+37 \cite{Milagro}, which is the brightest Milagro gamma-ray source in the Cygnus region. The measured flux was about 80\% of the Crab Nebula flux above 12 TeV and with no identified counterparts at other wavelengths. Its extent was measured to be about 1$\deg$ with the most plausable interpretation that the PWN is produced by the young, energetic pulsar PSR J2021+3651 ($\dot{E}=3.4 \times 10^{36}$ erg/s). 

Previous short observations with VERITAS \cite{VER_nodetection} and MAGIC \cite{Bartko} in 2008 led to upper limits consistent with the Milagro source being extended and having a hard spectral energy distribution. The ARGO-YBJ experiment also reported a non-detection of the source \cite{Bartoli}, with a claim that the source may be variable. However the Tibet Air Shower array has detected an extended VHE source in the same direction \cite{Wang}.

In 2014, the VERITAS collaboration performed a deep observation of this region \cite{VER-14}. VERITAS resolved MGRO J2019+37 into a point source VER J2016+371 and an extended source VER J2019+368, claiming the latter is the cause for the majority of Milagro emission. 
Its centroid is located at $\alpha_{J 2000}=20^{h} 19^{m} 25^{s} \pm 72_{s t a t}^{s}, \delta_{J 2000}=36^{\circ} 48^{\prime} 14^{\prime \prime} \pm 58^{\prime \prime}$ with an angular extension estimated to be $0^{\circ} .34 \pm 0^{\circ} .03_{s t a t}$ along the major axis and $0^{\circ} .13 \pm 0^{\circ} .02_{s t a t}$ along the minor axis with the
orientation angle 71$\deg$ east of north.
The energy spectrum of VER J2019+368 was derived from a circular region of 0.5$\deg$ radius and extends from 1 to 30 TeV where it was fit by a power law model. The photon index was determined to be $\gamma=1.75 \pm 0.08_{stat} \pm 0.2_{sys}$ with a differential flux at 5 TeV of $\left(8.1 \pm 0.7_{stat} \pm 1.6_{sys}\right) \times 10^{-14} \mathrm{TeV}^{-1} \mathrm{cm}^{-2} \mathrm{s}^{-1}$.

In the 2017 survey \cite{HAWC} HAWC published the data for 2HWC J2019+367. It has a spectral index of $\gamma=-2.29\pm0.06$ with a differential flux at 7 TeV of $30.2 \pm 3.1 \times 10^{-15} \mathrm{TeV}^{-1} \mathrm{cm}^{-2} \mathrm{s}^{-1}$. The integrated flux obtained by the fit of the HAWC source is more consistent with the Milagro measurement than the VERITAS measurement.

VERITAS again performed a study of this area in 2018 \cite{VER-18}. The source was fit with an asymmetric Gaussian with semi-major axis of $0.34 \pm 0.02_{ stat } \pm 0.01_{s y s} $ degrees and semi-minor axis of $0.14 \pm 0.01_{\ stat } \pm 0.02_{ sys }$ degrees. However the spectrum has significantly lower flux compared to the previous VERITAS spectrum \cite{VER-14} due to the smaller region used in the recent analysis. Also the spectral index was not as hard as in previous measurements. The analysis resulted in the following power law fit:
$N_{0}=\left(1.02 \pm 0.11_{s t a t} \pm 0.20_{s v s}\right) \times \mathrm10^{-16}\mathrm{GeV^{-1}cm^{-2}s^{-1}}$, $ E_{0}=3110$ GeV$, \gamma=1.98 \pm 0.09_{ stat } \pm 0.20_{s y s}$.

The aim of MAGIC observations presented here is to perform measurements in the $\gamma$-ray domain above several tens of TeV, which has not been deeply studied so far, expecting to find the hard spectrum extending above tens of TeV.
Since pulsar wind nebulae (PWN) are among the possible galactic sources of cosmic rays, it is very interesting to try to measure the maximum energy of the emission particles for this source.

\section{Very Large Zenith Angle Observations}
Due to the extremely low count rates at these energies, the detection of even the brightest sources requires either very long observations or extremely large detector arrays (like the future Cherenkov Telescope Array). 
The MAGIC Collaboration recently developed a new method to enable observations at Very Large Zenith Angles (VLZA) \cite{NIMA}, reaching $\sim$ $km^2$ collection areas by observing above $\sim$60$\deg$ zenith.
VLZA observations at the same time lead to a rise of the low energy threshold to several TeV, rendering the telescope sensitive only to the highest energy gamma rays.
The performance of the instrument during these observations depends on several factors that need to be analyzed such as the quickly changing atmosphere thickness with the very large zenith angle, large distance to the $\gamma$-ray induced air showers and the higher variability of the atmospheric transparency due to the increased atmospheric depth \cite{VLZA_ICRC}.

\section{MAGIC Observations and Datataking}
The MAGIC (Major Atmospheric Gamma Imaging Cherekov) telescopes are two 17m diameter Imaging Atmoshperic Cherenkov Telescopes (IACTs). They are located on the Canary Island of La Palma (Spain) at the Roque de los Muchachos Observatory at 2200 m above sea level.
Equipped with fast, 1039-pixel PMT cameras, the telescopes record images of extensive air showers (EAS) in stereoscopic mode, enabling the observation
of VHE gamma-ray sources \cite{MAGIC_hardware}.
When operated in standard mode the telescopes cover the energy range from
50 GeV to more than 50 TeV. The energy resolution is 16 \% and the angular resolution of the system is < 250 arcsec. The integral sensitivity in 50 h above 220 GeV is 0.66 \% of the flux of the Crab Nebula \cite{MAGIC_performance}.

The data sample presented here was accumulated from January 2018 until September 2018 in the zenith angle range 62$\deg$ - 80$\deg$ and comprises of $\simeq$ 45 hr of good-quality data 
taken during dark time. The observations were
performed in wobble-mode \cite{Fomin} at four symmetrical positions 0.4$\deg$ away from the desired sky position, allowing a simultaneous background estimation.
The pointing was centered on the PSR J2021+3651 location which is offset from the VER J2019+368 by about 0.3 deg. Because of this special care has been taken during the analysis as the source was not in the center of the skymaps.
For simulating the telescope performance diffuse Monte Carlo (MC) data was used. The MCs had been generated in a way that the simulated source is randomly distributed between 0 and 1.5$\deg$ from the camera center simulating an extended source.

\section{Data analysis}
The collected data has been analyzed with standard MAGIC Analysis and Reconstruction  Software (MARS) \cite{Zanin}\cite{MARS}. Events taken during bad weather conditions were discarded.
In this case, data taken during median of the aerosol transparency measurements 
lower than 80\% of that of an optimal night was discarded.
Standard MARS routines were used to reconstruct the incoming direction of the gamma rays.
A Random Forest (RF) multivariate analysis method was used to reconstruct the energy of the initiating particles of the extended air showers.
The cosmic ray background suppression was also based on the RF method.

A spatial likelihood analysis tool for MAGIC telescope data called Skyprism \cite{Skyprism} has been used  for extended source analysis.
The Skyprism analysis package enables a proper reconstruction of extended sources fluxes over the entire FoV, as the accuracy of the SkyPrism analysis does not degrade with the off-set from the telescope camera centre. 
Skyprism uses 2D spatial modeling of the measured sky signal. Applying the instrument response to an assumed source model, it can generate an image of the model as it would be seen by the telescope. This model, along with the estimated background map, is fitted to the measured sky image to estimate the most likely flux of the model sources in the chosen sky region.
The method chosen to reconstruct a background exposure model was the so called "Blind Map".
The blind map option calculates the combined background camera exposure using the median value in each pixel of the single wobble pointing normalized exposure times.
Typical selection cuts with 90\% efficiency, that were based on the MCs, for $\gamma$-ray/hadron separation were used.
The source was modeled as an extended source with power law spectrum and the model image was assumed to be an asymmetric 2D Gaussian. The 1$\sigma$ angular extension was set as $0.34\deg$ along the major axis, and $0.14\deg$ along the minor axis.
Background was chosen to be isotropic with a log parabola spectrum.

\section{Results} 




%

The source was detected with a significance exceeding 5$\sigma$ above 1 TeV.
The integrated flux obtained in the 1-30 TeV range is $(3.9\pm 0.6)\times10^{-12} \mathrm{cm^{-2}s^{-1}}$.
Result given here provides a higher integrated flux compared to the results published by VERITAS, where it was obtained to be $1.7\times10^{-12} \mathrm{cm^{-2}s^{-1}}$ in 2014 and $9.5\times10^{-13} \mathrm{cm^{-2}s^{-1}}$ in 2018.
This can be explained by the smaller integration region used by VERITAS in 2014 and 2018, where it was set to 0.5$\deg$ and 0.23$\deg$ respectively.
The result obtained here is more in line with the HAWC result with an extended source assumption (R=0.7$\deg$) which amounts to $3.6\times10^{-12} \mathrm{cm^{-2}s^{-1}}$. This is expected as the used integration radius of HAWC contains $\sim$95\% of the events of the 2D Gaussian used in this analysis.


VLZA observations have shown to be a powerful tool to unravel the highest energy emission of extended sources. In the future it is expected more sources in the multi TeV energy range could be analyzed using VLZA method.


\section*{Acknowledgements}
Full MAGIC acknowledgments: \href{https://magic.mpp.mpg.de/acknowledgments_ICRC2019/}{https://magic.mpp.mpg.de/acknowledgments\_ICRC2019/}

%

\end{document}